# Topological Superconductor $Bi_2Te_3$/$NbSe_2$ heterostructures

Jin-Peng Xu[1†], Canhua Liu[1†], Mei-Xiao Wang[1], Jianfeng Ge[1], Zhi-Long Liu[1], Xiaojun Yang[2], Yan Chen[3], Ying Liu[1,4], Zhu-An Xu[2], Chun-Lei Gao[1], Dong Qian[1], Fu-Chun Zhang[2,5], Qi-Kun Xue[6], and Jin-Feng Jia[1*]

[1]Key Laboratory of Artificial Structures and Quantum Control (Ministry of Education), Department of Physics and Astronomy, Shanghai Jiao Tong University, Shanghai 200240, China
[2]State Key Laboratory of Silicon Materials and Department of Physics, Zhejiang University, Hangzhou 310027, China
[3]Department of Physics, Fudan University, Shanghai 200433, China
[4]Department of Physics, Pennsylvania State University, University Park, PA 16802, USA
[5]Department of Physics, Hong Kong University, Hong Kong, China
[6]State Key Laboratory for Low-Dimensional Quantum Physics, Department of Physics, Tsinghua University, Beijing 100084, China

**Topological superconductors (TSCs) have a full gap in the bulk and gapless surface states consisting of Majorana fermions, which have potential applications in fault-tolerant topological quantum computation. Because TSCs are very rare in nature, an alternative way to study the TSC is to artificially introduce superconductivity into the surface states of a topological insulator (TI) through proximity effect (PE)[1-4]. Here we report the first experimental realization of the PE induced TSC in $Bi_2Te_3$/$NbSe_2$ thin films as demonstrated by the density of states probed using scanning tunneling microscope. We observe Abrikosov vortices and lower energy bound states on the surface of topological insulator and the dependence of superconducting coherence length on the film thickness and magnetic field, which are attributed to the superconductivity in the topological surface states. This work demonstrates the practical feasibility of fabricating a TSC with individual Majorana fermions inside superconducting vortex as predicted in theory and accomplishes the pre-requisite step towards searching for Majorana fermions in the PE induced TSCs.**

† J.P.X. and C. L. contributed equally to this work.
* Corresponding author: jfjia@sjtu.edu.cn.

When a normal metal (NM) and a superconductor (SC) are brought into contact, Cooper pairs will be introduced into the NM through the interface, resulting in a superconducting energy gap at the Fermi level ($E_F$) in NM. This phenomenon is called the superconducting PE[5]. Since the pairing symmetry of the PE induced Cooper pairs is determined by the electronic structures of the NM, TSCs that have p-wave-like pairing symmetry as well as time reversal symmetry can be obtained by the combination of a simple s-wave SC and a NM with a peculiar band structure that requires an odd number of spin non-degenerate electronic bands crossing $E_F$[3, 4]. TSCs are predicted to host Majorana fermions and thus attract a great deal of research interest. For example, a semiconductor nanowire with strong spin-orbit coupling fabricated on an s-wave SC may realize bound Majorana fermions located at the wire's two ends in an external magnetic field[6, 7]. Another proposed material for the construction of a TSC is a three dimensional TI that has a spin textured topological surface state band located in the bulk energy gap[8, 9]. Fu and Kane predicted that when Cooper pairs are introduced to its topological surface states via PE, the surface of the TI will turn into a TSC that can harbor Majorana fermions bound at Abrikosov vortex cores[2].

Recently, PE induced superconductivity in surface states of a TI has been manifested in electronic transport measurement on several TI/SC heterostructures[10-14], but the pairing symmetry has not been clarified in experiment yet. While an indirect signature of Majorana fermions was revealed in a Josephson current measurement performed on an Al/$Bi_2Se_3$/Al junction[12], more explicit evidence for the TSC with Majorana fermions can be obtained by direct detection of zero-energy bound states at Abrikosov vortex cores in a TI/SC heterostructure. Although Abrikosov vortices on ordinary SCs have been widely demonstrated in experiments[15-17], they are yet to be observed in a PE-induced SC, mainly due to the difficulty in preparing an atomically smooth interface and surface of the NM/SC heterostructure. In our previous work, we succeeded in constructing a TI/SC heterostructure by growing a $Bi_2Se_3$ thin film on a $NbSe_2$ single crystal, where the coexistence of Cooper pairs and topological surface states was demonstrated[18]. In this work, we report observation of PE superconductivity on another TI/SC heterostructure, $Bi_2Te_3$/$NbSe_2$, using scanning tunnelling microscopy/spectroscopy (STM/S) by greatly improving the film quality. We have systematically studied multi quantum-layers (QL) of $Bi_2Te_3$ on top of $NbSe_2$. When the Fermi level is within the bulk gap, the superconducting tunnelling spectra of the $Bi_2Te_3$/$NbSe_2$ surface deviate from the usual s-wave spectra, indicating PE superconductivity on the surface. We have succeeded in observing Abrikosov vortices and bound states inside the vortex core, in which a zero-bias bound state, or Majorana fermion has been predicted in theory. These experimental results provide the evidence for PE induced topological superconducting states on the surfaces of the TI films, which paves the way for detecting Majorana fermions -or even an individual Majorana fermion - on a TI/SC heterostructure.

The growth mode of $Bi_2Te_3$ is the same as that of $Bi_2Se_3$ on $NbSe_2$[18], i.e., layer by layer growth as shown in Figs. 1(a)-1(b). The excellent crystallization of the $Bi_2Te_3$ film is seen in its STM image with an atomic resolution. Due to its higher growth temperature, the $Bi_2Te_3$ film has very large terraces that are crucial for observing vortices in this work. The $Bi_2Te_3$ film is not interfaced with the $NbSe_2$ substrate directly but separated by a Bi layer in between, so that the first quintuple layer (QL) of $Bi_2Te_3$ has a step height of 1.6nm in STM images. All other QLs have the thickness of 1.0nm, similar to those grown on a Si(111) substrate[19]. Previous studies found that the topological surface state of a $Bi_2Te_3$ thin film grown on Si(111) does not form a Dirac cone until the thickness reaches 2 to 4QL because of the hybridization between the top and bottom surface states of the film[20-22]. Once the Dirac cone forms, the band structure near the Fermi energy does not change dramatically except for a rigid energy shift with increasing the thickness of the $Bi_2Te_3$ film. On the $NbSe_2$ substrate, we also observed a similar evolution of the electronic states in STS data with the increase of $Bi_2Te_3$ thickness, as shown in Fig. 1(c). The differential conductance curve (dI/dV spectrum) taken with STS, obtained on a 20QL $Bi_2Te_3$ film grown on a Si(111) surface has a deformed U-shape segment in the energy range between the bulk valence band maximum (VBM) and the conduction band minimum (CBM), as indicated by blue and red arrows in Fig. 1(c), respectively. Similar deformed U-shape segments are also seen in dI/dV spectra taken on $Bi_2Te_3$/$NbSe_2$ at thicknesses of no less than 3QL. With the increase of $Bi_2Te_3$ thickness, the deformed U-shape segment shifts to higher binding energy while keeping its size in energy scale. This indicates that the topological surface states on $Bi_2Te_3$/$NbSe_2$ come into existence at a thickness of 3QL, very similar behavior to that of $Bi_2Te_3$ films grown on Si(111) substrates[20].

The superconducting energy gap was observed on the $Bi_2Te_3$ thin films in STS data. Fig. 2(a) is a series of dI/dV spectra taken on $Bi_2Te_3$/$NbSe_2$ at 0.4K, showing strong thickness dependence. Fig. 2(b) presents the dI/dV spectra of bare $NbSe_2$, 2QL, and 3QL $Bi_2Te_3$ films. Up to a thickness of 2QL, the STS curve has a flat bottom that touches the zero value of the differential conductance at around zero bias. The STS curve can be well fitted by an s-wave BCS-type spectrum function [Fig. 2(b)-bottom and middle]. In contrast, the spectra of $Bi_2Te_3$ films of more than 2QL have a non-flat bottom that does not touch the zero of the differential conductance. Clearly, an s-wave BCS-type spectral function can no longer fit the spectrum of a 3QL $Bi_2Te_3$ film [Fig. 2(b)-top]. A simple s-wave BCS-like fitting curve can neither reproduce the spectrum near zero-bias, nor the sharp coherence peak near the position of the energy gap. The differences could be induced by PE, where pair field diminished into the films, therefore the coherence length increases, and also the quasiparticle lifetime increases so that states appear at bottom of gap. However, the similar behavior was not observed on 1 or 2QL films, and abruptly became obvious on 3QL films, where the topological surface states form. So, the formation of topological surface states might play some role for the deviation of the 3QL's dI/dV spectrum from s-wave BCS-like behavior[23, 24]. Note that a simple p-wave like pairing for topological superconductor gives essentially the same dI/dV spectra as s-wave BCS at the relevant energy region.

The data up to 11QL are fitted by thermally broadened s-wave BCS-like curves[25], and the results are summarized in Fig. 2(c). Roughly speaking, the gap value decreases exponentially as the thickness increases, qualitatively consistent with the energy gap decay for PE-induced superconductivity. However, a kink can obviously be seen at 6QL, where the Fermi level is already in the bulk band gap [see Fig.1(c)]. The unusual character implies two-band superconducting, i.e., bulk superconducting state and surface superconducting state. At 6QL or more, surface superconducting state may dominate, so the superconducting gap decays slower.

Since the formation of the small energy gap at the Fermi level is due to quasi particle excitations in a superconductor, the dI/dV spectra measured on $Bi_2Te_3$/$NbSe_2$ should also have strong dependence on external magnetic fields, which suppress the formation of Cooper pairs. In Fig. 3(a), we show a series of spatially averaged dI/dV spectra taken on a 3QL $Bi_2Te_3$/$NbSe_2$ sample at various perpendicular magnetic fields. With the increase of the magnetic field, the energy gap becomes smaller and gap feature near the zero bias in the dI/dV spectra becomes shallower. The energy gap disappears at about 2.4T, which is substantially smaller than the upper (perpendicular) critical field $H_{c2}$ (3.2 T) of the $NbSe_2$ substrate.

Large terraces of the $Bi_2Te_3$/$NbSe_2$ surface make it possible to image Abrikosov vortices with STS[26]. Figs. 3(b)-3(c) show dI/dV maps at zero bias, i.e., the contour of zero bias conductance (ZBC), recorded on a 3QL $Bi_2Te_3$/$NbSe_2$ and bare $NbSe_2$ surfaces under perpendicular magnetic fields. It is seen from Fig. 3(c) that the vortices exhibit a highly ordered hexagonal lattice just like those observed on the clean $NbSe_2$ surface shown in Fig. 3(b). This is the first time the vortex has been clearly observed in a proximity induced topological superconductor. Due to the crystalline band structure and the interaction of the neighboring vortices in the hexagonal lattice, a six-fold symmetry is explicitly observed in the vortex images of the bare $NbSe_2$ surface [Fig. 3(d)]. The same symmetry is also present on the ZBC contour of the 5QL $Bi_2Te_3$ film [Fig. 3(e)]. The growth of $Bi_2Te_3$ films on $NbSe_2$ does not change the orientation of the vortex lattice to avoid extra energy consumption for magnetic flux penetrating the Bi$_2$Te$_3$/NbSe$_2$ samples. This may be seen with the aid of the two hexagons superimposed on Figs. 3(d) and 3(e) (dashed lines). Also, from the size of the unit cell of the vortex lattice, we can calculate the magnetic flux penetrating through one vortex cylinder. The obtained value is very close to a magnetic flux quantum, i.e., $\Phi_0 = h/2e$, in which *h* is Planck's constant and *e* is the electric charge of an electron.

By carefully comparing the vortices obtained on NbSe$_2$ and 3QL Bi$_2$Te$_3$ shown in Figs. 3(b) and 3(c), respectively, one can see that the vortex size is a little smaller on NbSe$_2$ than that on 3QL Bi$_2$Te$_3$. We investigated the spatial extension of the vortex for different Bi$_2$Te$_3$ thickness. The ZBC line-profile crossing through the center of the vortex can be very well fitted by the formula below, derived from the Ginzburg-Landau (GL) expression for the superconducting order parameter[17]:

$$\sigma(r,0) = \sigma_0 + (1 - \sigma_0) \times \{1 - \tanh[r/(\sqrt{2}\xi)]\}, \quad (1)$$

where $\sigma_0$ is the normalized ZBC away from a vortex core, *r* is the distance to the vortex center and $\xi$ is the GL coherence length in plane. The experimental data and fitted results for bare NbSe$_2$ and 3QL Bi$_2$Te$_3$/NbSe$_2$ are shown in Figs. 4(a), giving $\xi_{NbSe_2}$=16nm and $\xi_{5QL}$=29nm at 0.4K and 0.1T. Similar analyses were also performed on other samples, finding a monotonic increase of coherence length with Bi$_2$Te$_3$ thickness as shown in Fig. 4(b). This is consistent with the above result that $H_{c2}$ of 3QL Bi$_2$Te$_3$/NbSe$_2$ is smaller than that of bare NbSe$_2$, since a larger $\xi$ gives a smaller $H_{c2}$ according to the GL expression, $H_{c2} = \Phi_0/(2\pi\xi^2)$.

As the in-plane $\xi$ of a NbSe$_2$ single crystal varies from 7.2nm to 28.2nm in previous reports[27, 28], $\xi$ obtained above for the bare NbSe$_2$ is a reasonable value. For Bi$_2$Te$_3$ films, the Pippard's coherence length can be estimated to be about 116nm by the formula $\xi_0 = hv_F/(\pi^2\Delta)$, in which $v_F$ is the Fermi velocity, previously reported to be 3.32x10$^5$m/s[24]. In the dirty limit, which is the case for the NbSe$_2$ substrate and thus should be also for the proximity induced TI/SC heterostructure, $\xi = \xi_0$ at T=0 K. The discrepancy between the estimation and the above experimental results lies in the fact that the $Bi_2Te_3$ film is a PE-induced superconductor, and the properties of the induced Cooper pairs are inevitably influenced by the parent superconductor of $NbSe_2$. As the thickness of the $Bi_2Te_3$ film increases, the influence from the $NbSe_2$ substrate on the $Bi_2Te_3$ film becomes weaker, resulting in a longer coherence length.

Variation of magnetic field leads to changes of the vortex size and the coherence length. The magnetic field dependence of the coherence length is shown in Fig. 4(c) for 5QL $Bi_2Te_3$/$NbSe_2$. As magnetic field increases, the coherence length decreases initially and then saturates at magnetic field reaches about 0.7T. For a single band s-wave superconductor, the vortex size or the coherence length is insensitive to the magnetic field at the weak field. The strong dependence of the vortex size with the field is a clear evidence of non-single band s-wave nature for the vortex state and is consistent with the scenario that both the bulk conduction and surface states are superconducting and contributing to the vortex state. In brief, all the facts in our experiment demonstrate the topological surface states on $Bi_2Te_3$/$NbSe_2$ are superconducting and $Bi_2Te_3$/$NbSe_2$ is a topological superconductor, which can host the Majorana fermions in the vortex core. It is noted that the superconductivity in topological surface states has been confirmed recently by ARPES[29].

Majorana mode is the zero-energy bound state in the vortex core, and the first step to detect it is to observe the peak caused by the bound states in the vortex core. The bound-state peaks were observed in 1-6QL $Bi_2Te_3$/$NbSe_2$ as shown in Fig. 4(d). However, these peaks contain all bound states near the Fermi energy, could not be identified as a signature of Majorana fermions. Actually, the peaks are also observed on bare $NbSe_2$ and 1-2QL $Bi_2Te_3$ films, and their heights are even higher than those of 3-6QL $Bi_2Te_3$ films. The bound states occur at energies where constructive interference of multiply reflected electron-like and hole-like states forms. In a 2D chiral p-wave SC, the energy separation of the bound states is theoretically predicted to be $\delta E = \Delta_0^2 / E_F$[30], which is the minigap protecting the zero-energy Majorana fermion excitations from thermal effects. In recent theoretical work[31], the minigap of a PE-induced TSC is evaluated by $\delta E \approx 0.83\Delta_0^2/\sqrt{\Delta_0^2 + E_F^2}$. In order to increase the robustness of Majorana fermions to thermal effect and to resolve them with STS, one needs to decrease the sample temperature as much as possible and increase $\delta E$ by shifting $E_F$ towards the Dirac point. For a 5QL $Bi_2Te_3$/$NbSe_2$ sample, for example, at a temperature of 300mK that corresponds to a thermal energy of 0.026meV, if $E_F$ can be shifted to within 5meV of the Dirac point, the $\delta E$ will be about 0.1meV, which is much larger than the thermo energy and can be resolved by STS. The $E_F$ can be tuned by doping, or alloying $Sb_2Te_3$ and $Bi_2Te_3$[32], etc., However, chemical doping or engineering will reduce the sample quality and thus requires a much greater experimental efforts to achieve the precise control of the Fermi level, and tuning $E_F$ in the gap also suppresses the PE, resulting a smaller $T_c$, which makes the experiment more challenging.

In summary, we have provided experimental evidences for the superconductivity in topological surface states induced by PE. Our work confirms the theoretical prediction that TSC can be realized in a TI/SC heterostructure. Abrikosov vortex lattices and bound states at the vortex cores have been successfully observed for the first time on the PE induced artificial topological superconducting $Bi_2Te_3$/$NbSe_2$ heterostructure. Our study is an important step in search predicted Majorana fermions inside the vortex core of TSC.

**Methods**

All experiments were performed in situ on a commercial apparatus with base pressures of $3\times10^{-10}$torr for sample growth and $7\times10^{-11}$torr for STM measurement. $NbSe_2$ crystal was cleaved at room temperature in UHV after sufficient degassing at 250°C. Te and Bi atoms were co-deposited onto the $NbSe_2$ surface at 250°C. 5 minutes post annealing was conducted to eliminate excess Te atoms in the $Bi_2Te_3$ films. All prepared samples were transferred to a cooling stage kept at 4.2K for STM measurement, in which electrochemically etched tungsten tips were used after heating and silver-decoration in situ. A lower sample temperature of 0.4K was achieved by using liquid 3He. To obtain dI/dV spectra at a given location, the tip-sample separation was held constant and a lock-in amplifier was used to modulate the bias voltage by dV (0.15mV or 3mV depending on the spectral range of interest) with a frequency of 991Hz. The vortex image is the record of each pixel's ZBC while bias voltage is ramped from 0.05mV to 0V with feedback off.

Acknowledgements
This work was supported by the National Basic Research Program of China (Grant No. 2012CB927401, 2011CB921902, 2013CB921902, 2011CB922200, 2012CB921604), NSFC (Grant No. 91021002, 11174199, 11134008, 11274228, 11074043, 11274269), Shanghai Committee of Science and Technology, China (No. 11JC1405000, 11PJ1405200, 12JC1405300), Shanghai Municipal Education Commission (11ZZ17).

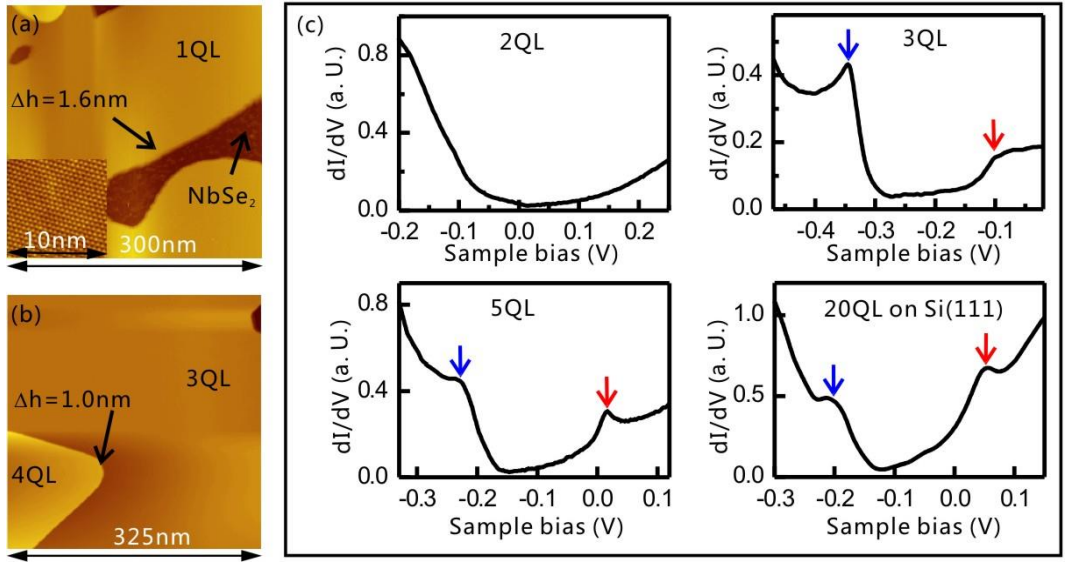

Fig. 1. Morphology and electronic density of states of $Bi_2Te_3$ thin films of different coverages, grown on a $NbSe_2$ substrate. (a) and (b) Large-scale STM images of $Bi_2Te_3$ thin films. The insert in (a) shows the atomic resolution of the $Bi_2Te_3$ surface taken on a 3QL terrace. (c) dI/dV spectra measured at 4.2K on 2QL, 3QL and 5QL $Bi_2Te_3/NbSe_2$ heterostructures and on a 20QL $Bi_2Te_3$ film grown on a Si(111) surface. Blue and Red arrows indicate the energy position of VBM and CBM, respectively.

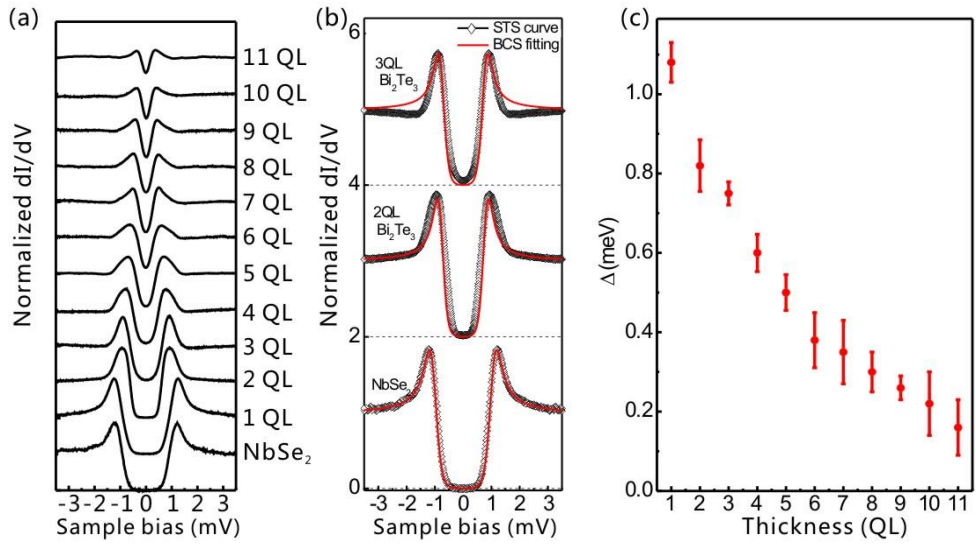

Fig. 2. Superconducting energy gap observed on $Bi_2Te_3$/$NbSe_2$. (a) A series of dI/dV spectra taken on different thicknesses of $Bi_2Te_3$ thin films at 0.4K. (b) dI/dV spectra measured at 0.4K of bare $NbSe_2$, and 2QL and 3QL of $Bi_2Te_3$ thin films all superimposed with standard BCS-like fitting results. The latter two spectra are shifted upward by 2 and 4, respectively. (c) Thickness dependence of the energy gap obtained from the BCS-like fitting.

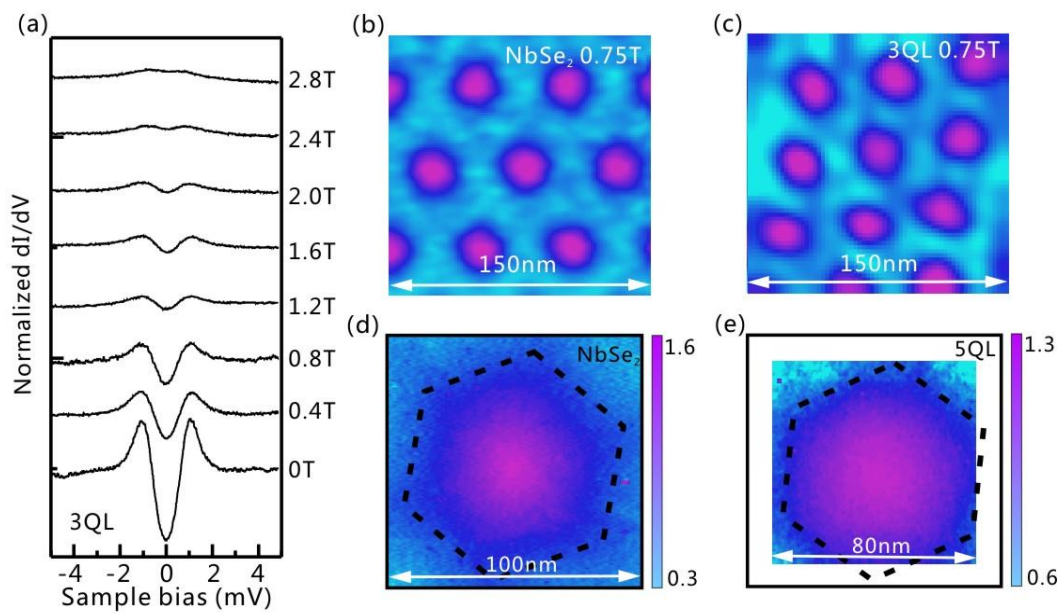

Fig. 3 (a) Dependence of dI/dV spectra on magnetic field measured on a 3QL $Bi_2Te_3$ thin film at 0.4K. (b) and (c) Zero bias dI/dV maps measured at 0.4K and 0.75T for $NbSe_2$ and 3QL $Bi_2Te_3/NbSe_2$ heterostructures. (d) and (e) Zero bias dI/dV maps for a single vortex measured at 0.4K and 0.1T on $NbSe_2$ and 5QL $Bi_2Te_3/NbSe_2$. The superimposed hexagons in dashed lines indicate the shape of the vortices.

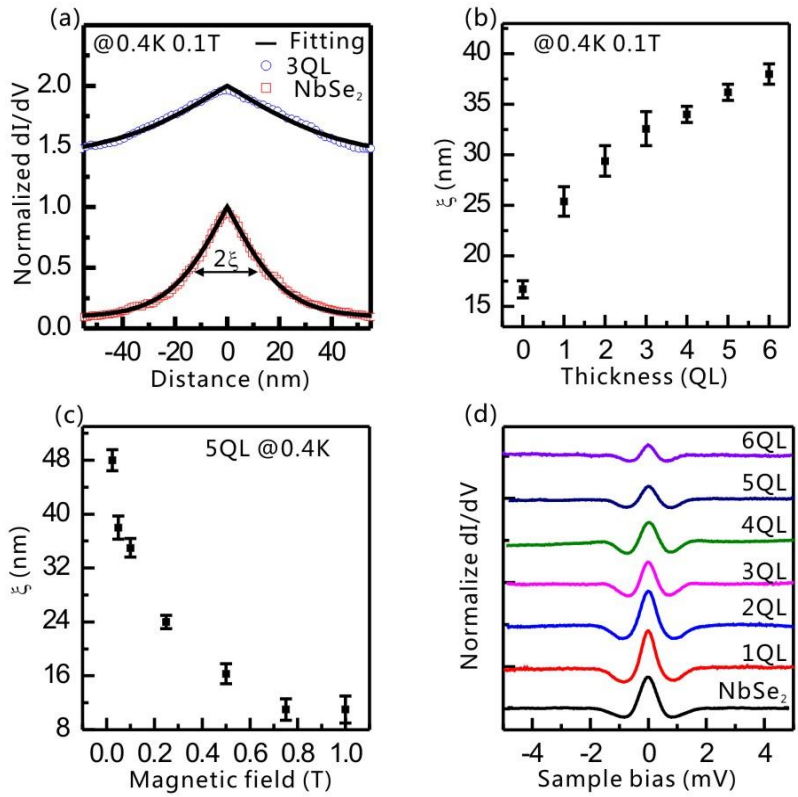

Fig. 4 (a) Normalized ZBC profiles crossing through the centers of vortices at 0.4K and 0.1T on NbSe$_2$ and 3QL Bi$_2$Te$_3$, respectively. The superimposed lines are fitted results using Eq. (1). The 3QL data is shifted upward by 1.5 for a better view. The obtained coherence length as a function of thickness is summarized in (b). (c) The coherence length as a function of magnetic field measured on 5QL Bi$_2$Te$_3$/NbSe$_2$. (d) Thickness dependence of dI/dV spectra measured at centers of vortices on NbSe$_2$ and 1-6QL Bi$_2$Te$_3$/NbSe$_2$ at 0.4 K.